\documentclass[%
 aip,
 amsmath,amssymb,
 reprint,%
]{revtex4-1}

\usepackage{graphicx}
\usepackage{dcolumn}
\usepackage{bm}

\usepackage[utf8]{inputenc}
\usepackage[T1]{fontenc}
\usepackage{mathptmx}
\usepackage{etoolbox}
\usepackage[colorlinks,citecolor=blue]{hyperref}
\makeatletter
\def\@email#1#2{%
 \endgroup
 \patchcmd{\titleblock@produce}
  {\frontmatter@RRAPformat}
  {\frontmatter@RRAPformat{\produce@RRAP{*#1\href{mailto:#2}{#2}}}\frontmatter@RRAPformat}
  {}{}
}%
\makeatother
\begin{document}

\preprint{AIP/123-QED}

\title{Effect of Positive Polarity in an Inertial Electrostatic Confinement Fusion Device: Electron Confinement, X-ray Production, and Radiography}
\author{D. Bhattacharjee}

\author{S. R. Mohanty}
\altaffiliation[Also at:]{Homi Bhabha National Institute, Anushaktinagar, Mumbai, Maharashtra, 400094, India; Email:smruti@cppipr.res.in}
\affiliation{Centre of Plasma Physics-Institute for Plasma Research,Sonapur,Kamrup-782402,India}

\author{S. Adhikari}
\affiliation{Department of Physics, University of Oslo, PO Box 1048 Blindern, NO-0316 Oslo, Norway}

\date{\today}

\begin{abstract}
The conventional inertial electrostatic confinement fusion (IECF) operation is based on the application of high negative voltage to the central grid which results in the production of neutrons due to the fusion of lighter ions. The neutron has enormous applications in diversified fields. Apart from the neutrons, it can also be used as an application based x-ray source by altering the polarity of the central grid. In this work, the electron dynamics during the positive polarity of the central grid have been studied using an object-oriented particle-in-cell code (XOOPIC). The simulated trapped electron density inside the anode is found to be of the order of $10^{16}$ m$^{-3}$ when $10$ kV is applied to the anode. The re-circulatory characteristics of the electrons are also studied from the velocity distribution function. The x-ray production, imaging and radiography have been investigated at different voltages and using different structure of the anode. The x-ray emitting zone have been studied via pinhole imaging technique. Lastly, the radiography of metallic as well as biological samples have been studied in the later part of this paper. This study shows the versatile nature of the IECF device in terms of its applications, both in the field of neutron and x-ray.
\end{abstract}

\maketitle

\section{\label{sec:1}Introduction:}
The confinement of plasma particles to produce nuclear fusion for the purpose of thermonuclear energy is one of the most demanding topics in the current scenario. Magnetic confinement fusion (MCF), inertial confinement fusion (ICF) are some of the commonly used plasma confinement techniques. The inertial electrostatic confinement fusion (IECF) is an alternate technique in which the confinement of plasma particles takes place due to the application of a purely electrostatic field between partially transparent gridded electrodes. The simplicity, less complex design, compactness and portability make the IECF system a low cost fusion reactor and neutron generator. The basic advantage of the device is its applicability in numerous fields, such as in the production of radioisotopes\cite{weidner2003production,cipiti2005production}, in landmine detection\cite{sorebo2009special,yoshikawa2007research,yoshikawa2009research}, in the development of plasma jet for rocket propulsion\cite{bussard1995inertial}, and in many other fields\cite{miley2014inertial,kulcinski2009near}. The IECF device basically consists of one or more transparent gridded assembly concentrically placed inside a spherical or cylindrical chamber. The innermost grid serves as the cathode in which high negative voltage is supplied. Breakdown of the fuel gas (deuterium, tritium, $^3$He, etc.) takes place between the electrodes and the ions re-circulate across the grid openings with very high energy depending upon the applied voltage. The oscillating ions get trapped inside the potential well and form a high density core region inside the cathode. Finally, the beam-beam or beam-background collision initiates the fusion process as a result of which neutrons are produced\cite{miley2014inertial,thorson1997convergence,thorson1998fusion}. Since the development of the very first IECF device\cite{hirsch1967inertial,farnsworth1966electric}, the researchers carried out geometrical and technical modification of the device in order to increase the neutron yield\cite{tomiyasu2009effects,miley1998accelerator,miley1997iec,rasmussen2020characterization,ohnishi1998study,nadler1992characterization}, over the years. Recently, our team of researchers at CPP-IPR have carried out experiments in the cylindrical IECF device with some success in neutron production and its application in explosive detection\cite{buzarbaruah2015design,buzarbaruah2017study,buzarbaruah2018study,bhattacharjee2019studies,bhattacharjee2021neutron}. \par Apart from the conventional way of generating neutrons via negative biasing of the central grid, the IECF device has a basic application when the polarity of the central grid is reversed, i.e., positively biased. In such case, the electrons are the ones re-circulating across the central grid and the x-ray is the basic product due to the electron-target interaction. The same device can be used as a neutron generator when the central electrode is negatively biased and as an x-ray generator when the electrode biasing is reversed. Miley and his group carried out the x-ray emission experiments via positive biasing of the central grid and adding electron emitters inside a spherical chamber. They have termed the device as a tuneable x-ray source\cite{miley1999portable,gu1995spherical}. Pulsed x-ray has been produced in a table-top cylindrical IECF device by another group\cite{elaragi2018operation}. In some experiments, both neutron and x-ray has been produced during negative polarity of the grid in which the neutrons are produced from the central region and the x-ray from the wall region of the chamber\cite{rasmussen2020characterization,bhattacharjee2021neutron}. However, due to the broad x-ray emitting zone the x-ray flux is less in such cases and hence, it has limitations as far as x-ray applications are concerned. \par In this work, x-ray has been produced via positive biasing of the central electrode in the cylindrical IECF device in a steady-state operation. Unlike the pulsed operation, the steady-state allows the operator to regulate the voltage and current to customize the inspections in order to accommodate the needs. In other words, one can tune the voltage and current in the steady-state x-ray according to the nature of the sample through which it passes. Four filaments are also used as electron emitters to breakdown the neutral hydrogen gas in order to produce glow discharge plasma. The electron dynamics due to the positive biasing of the central grid in the device is studied computationally by using an open source object-oriented particle-in-cell (XOOPIC) code. The produced x-ray is used for pinhole imaging and radiography of some metallic as well as biological samples. The next section (\ref{sec:2}) of this paper contains the modeling of the device to carry out the simulation. Section (\ref{sec:3}) explains the experimental set up and procedure of this work. Section (\ref{sec:4}) explains the obtained results including simulation, x-ray production, x-ray imaging and radiography. Finally, the last section of the paper contains the concluding remarks of this work.

\section{\label{sec:2}Modeling}
Electrostatic particle-in-cell (PIC) simulation have been carried out using XOOPIC code\cite{verboncoeur1995object,bhattacharjee2020kinetic} to characterise the particle behaviour inside the cylindrical IECF device. The code has the ability to operate in two-dimensional space in both Cartesian and cylindrical coordinate systems, including three-dimensional velocity space. In order to recreate the exact experimental conditions, the dimension of the simulation domain including grids and all the parameters are considered accordingly. A horizontal $2$D cross-section of the device along with the cross-section of the anode grids are considered as the simulation geometry [fig.(\ref{fig:2})]. Four electron emitters are also modelled near the four walls inside the simulation domain from which equal flux of electrons are emitted continuously. These primary electrons interact with the neutral gas (hydrogen) and produce positive hydrogen ions in the process through ionisation collision. A high positive potential is applied to the central anode grid as a result of which the electrons accelerate towards the grid. The time step ($\Delta t$) for the simulation has been chosen in order to satisfy the Courant condition\cite{de2013courant} and it is resolved from the fastest particle (electrons) present in the system\cite{birdsall2004plasma}:
\begin{figure}
	\centering
	\includegraphics[width=7cm]{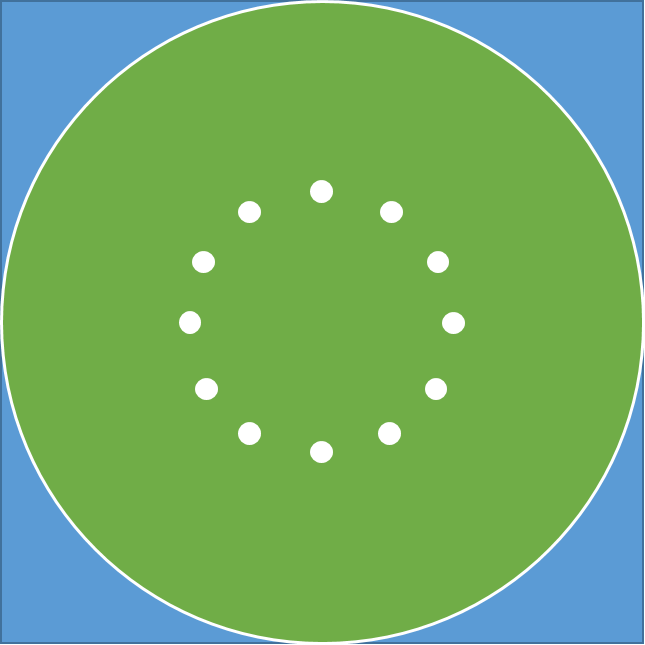}%
	\caption{Cross-section of the simulation domain. The white dots indicate the cross-section of the grid wires.  \label{fig:2}}%
\end{figure}
\begin{equation}
    \Delta t=0.3\times\frac{d}{v_{drift}} ~;~~d = \frac{1}{\sqrt{(\frac{1}{\Delta x^{2}}+\frac{1}{\Delta y^{2}})}} \label{eq1}
\end{equation}
Here, $v_{drift}$ is the drift velocity of the electrons, $\Delta x$ and $\Delta y$ are the cell sizes in the $x$ and $y$ directions, respectively. Putting appropriate values in the above equation, the $\Delta t$ is found to be of the order of $10^{-11}~s$. The size of the cell $\Delta x$ is considered in such a way that $\Delta x \leq \lambda_{D}$, where, $\lambda_{D}$ is the electron Debye length. XOOPIC also includes a Monte Carlo collision model which can incorporate elastic, ionization, excitation, and charge exchange collisions\cite{becker2004non}. Multigrid Poisson solver is used which describes the boundary conditions of the simulation\cite{becker2004non}. Considering all these parameters and conditions, the input file of the XOOPIC is developed for the simulation. In order to get a good resolution of phase space, a large number of macro-particles with relatively shorter domain length is appropriate. The simulation parameters are displayed in the table(\ref{Table 1}). The diagnostics are saved as ASCII file after the simulation and a few scripts are also developed in MATLAB\textsuperscript{\textregistered} to analyse the simulation data. 
\begin{table}
			\caption{Simulation parameters} 
			\label{Table 1}
			\begin{center}
			\footnotesize
                \begin{tabular}{@{}ll}
					\textbf{Parameters} & \textbf{Values} \\
					Grid size & $512\times512$ \\ 
					Length & $0.21$ m \\  
					Width & $0.21$~m \\ 
					Time step ($\Delta t$) & $10^{-11}$ s \\
					Specific weight & $10^{8}$ \\
					Background gas & Hydrogen \\
					Grid potential & $10$~kV \\
				\end{tabular} 
			\end{center}
		\end{table}%

\section{\label{sec:3}Experimental procedure:}
The cylindrical IECF device is made up of stainless steel having a diameter of $52$ cm and a height of $30$ cm. The inner electrode consists of a tungsten gridded assembly having multiple number of grid wires. During the conventional IECF operation, the central grid serves as the cathode in which high negative potential is applied in order to accelerate the ions to fusion relevant energies\cite{miley2014inertial}. Confinement of the ions is the basic criteria to exhibit fusion and produce particle, such as neutrons, protons, etc. However, in this particular study, the central electrode acts as the anode (positively biased) and the other electrode is the chamber wall itself which is grounded. A ceramic feedthrough is accessed through the topmost port of the chamber in order to provide high voltage to the grid assembly. The side wall of the chamber consists of multiple ports for gas inlet and outlet, viewing windows, filaments, pressure gauges, and to access diagnostic tools for plasma characterisation. The chamber is evacuated by using a rotary pump and a turbo molecular pump. The pressure inside the chamber is monitored in a display control unit (DCU) which is connected to a full range pressure gauge attached to the chamber. The hydrogen gas has been supplied into the chamber and a gas pressure of the order of $10^{-3}$ Torr is maintained during the experiments depending upon the required operating voltage.
\begin{figure}
	\centering
	\includegraphics[width=8cm]{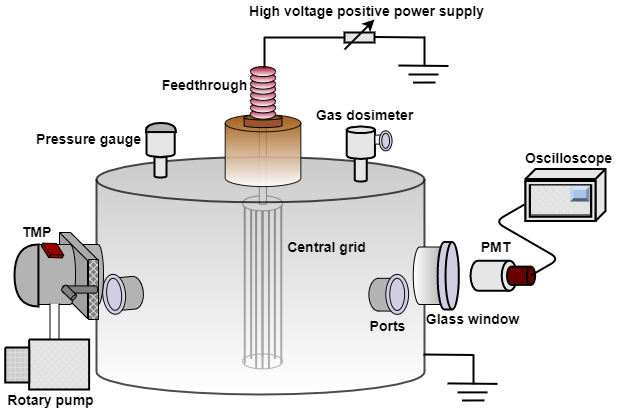}%
	\caption{Schematic diagram of the cylindrical IECF device with accessories. Here, PMT: photomultiplier tube, TMP: turbo molecular pump. \label{fig:1}}%
\end{figure}
\par In order to detect X-ray, we have placed the photomultiplier tube (PMT) near the device and it is connected to the oscilloscope to obtain the x-ray energy spectrum. At first, we have calibrated the channels of the PMT with a known source before obtaining the x-ray spectrum. Again, x-ray imaging have been performed by constructing a pinhole camera having a $200$ $\mu$m diameter of the pinhole. The pinhole camera is made up of a Cu assembly the thickness of which is large enough so that even very high energetic x-ray photon unable to pass through it. The x-ray photographic film (dental film having a dimension of $3.05\times4.05~\text{cm}^{2}$) is movable, as par requirement, inside the pinhole assembly. The pinhole camera along with the film is placed before the glass window to obtain the inverted image of the x-ray source. For the x-ray radiography purpose, we have attached the x-ray photographic film on the glass window from the outside. Contact radiography is basically performed in which the sample is attached to the film. After performing the experiments and having exposed to the x-ray, the film is processed in the developer and fixer solutions in order to obtain the permanent viewable image.

\section{\label{sec:4}Results and discussion:}
As already mentioned, a whole lot of experimental and theoretical works on ion dynamics and neutron production in the IECF device has been carried out by many researchers. However, in the reverse polarity of the central grid (positively biased) the electron dynamics plays a vital role as far as the x-ray production and radiography results are concerned. Therefore, we have discussed the electron behavior in the next subsection, briefly, and then x-ray generation and the results of radiography has been explained in the later parts.

\subsection{\label{sec:4.a}Electron dynamics in the IECF device:}
Electron dynamics is equally important as that of the ion in the fusion process inside the cathode during negative polarity of the central grid in the IECF device. In that case, the ions re-circulate along certain channels across the cathode grids and the electrons move in the opposite direction towards the wall of the chamber. Confinement of secondary electrons also takes place inside the potential well along-with the ions which results in the formation of multiple virtual electrodes inside the cathode\cite{bhattacharjee2020kinetic}. However, if the polarity of the central grid is altered, i.e., the grid is biased with positive voltage then, this time the electrons oscillate across the grid openings and the ions move towards the wall region. The oscillating frequency of the electrons, in this case, is much higher than that of the ions, in the previous case, due to the higher mobility of the electrons. \par
\begin{figure}
	\centering
	\includegraphics[width=8cm]{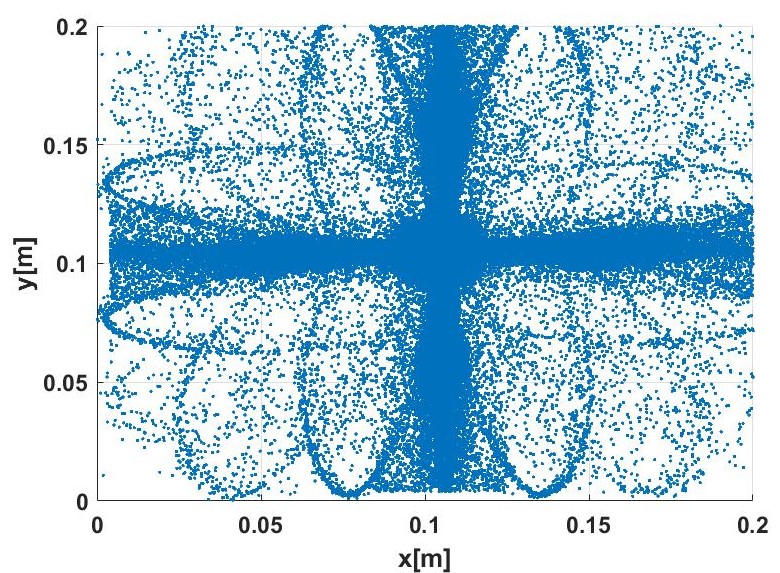}%
	\caption{Phase space plot of electrons during $10$ kV simulation. \label{fig:3}}%
\end{figure}

 In the simulation model we have applied a potential of $10$ kV to the anode grids and run it until the steady-state is reached. The electrons from the emitters accelerate towards the anode grids and re-circulate along the grid openings. The re-circulation process of the electrons can be visualized during the run-time of the simulation. Fig.(\ref{fig:3}) shows the phase space plot of the electrons after achieving the steady-state. The path followed by the electrons during their motion can be observed in this phase space. Apart from the straight line path of the electron beams originating from the emitters, they also follow the curved paths along the anode openings during the re-circulatory motion.
\begin{figure}
	\centering
	\includegraphics[width=8cm]{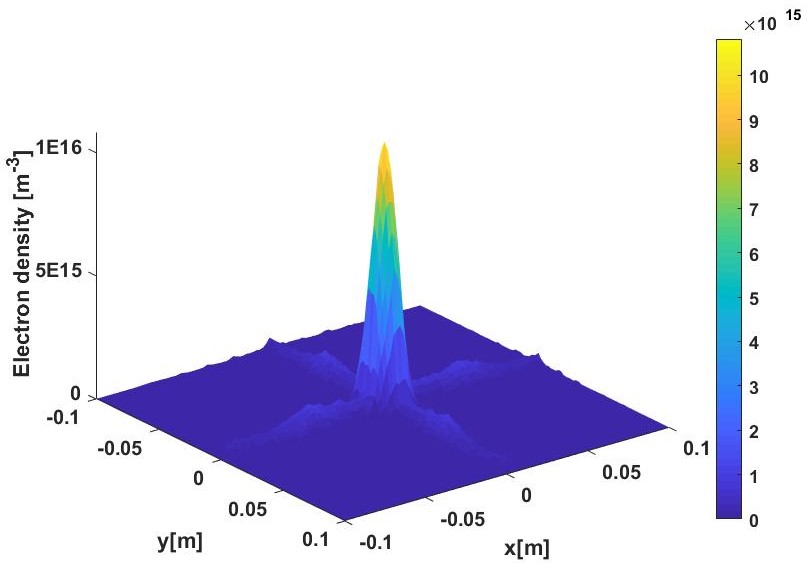}%
	\caption{Surface plot of the electron density profile at $10$ kV. Here, ($0$, $0$) coordinate represents the center of the device. \label{fig:4}}%
\end{figure}
The corresponding surface plot of electron density is shown in fig.(\ref{fig:4}). It clearly shows that the electrons are confined at the central region inside the anode and the peak density is found to be of the order of $10^{16}$ m$^{-3}$ at $10$ kV operation. This electron cloud density is found to be increasing with the applied potential, however, the order remains the same. The ion density is also found to be higher inside the grid due to the higher concentration of electron and higher ionisation rate inside the anode. \par In order to determine the velocity distribution of the electrons we have chosen three different location inside the simulation domain. One location is near the wall or emitter, the second one is in-between the emitter and the anode grids and the last one is inside the anode grids (central region). A function tool is used (in MATLAB) based on the normal kernal function which is a non-parametric representation of the probability density function (PDF) of the particle velocity. Similar approach was considered elsewhere\cite{schwager1990collector,adhikari2018ion}.
\begin{figure}
	\centering
	\includegraphics[width=8cm]{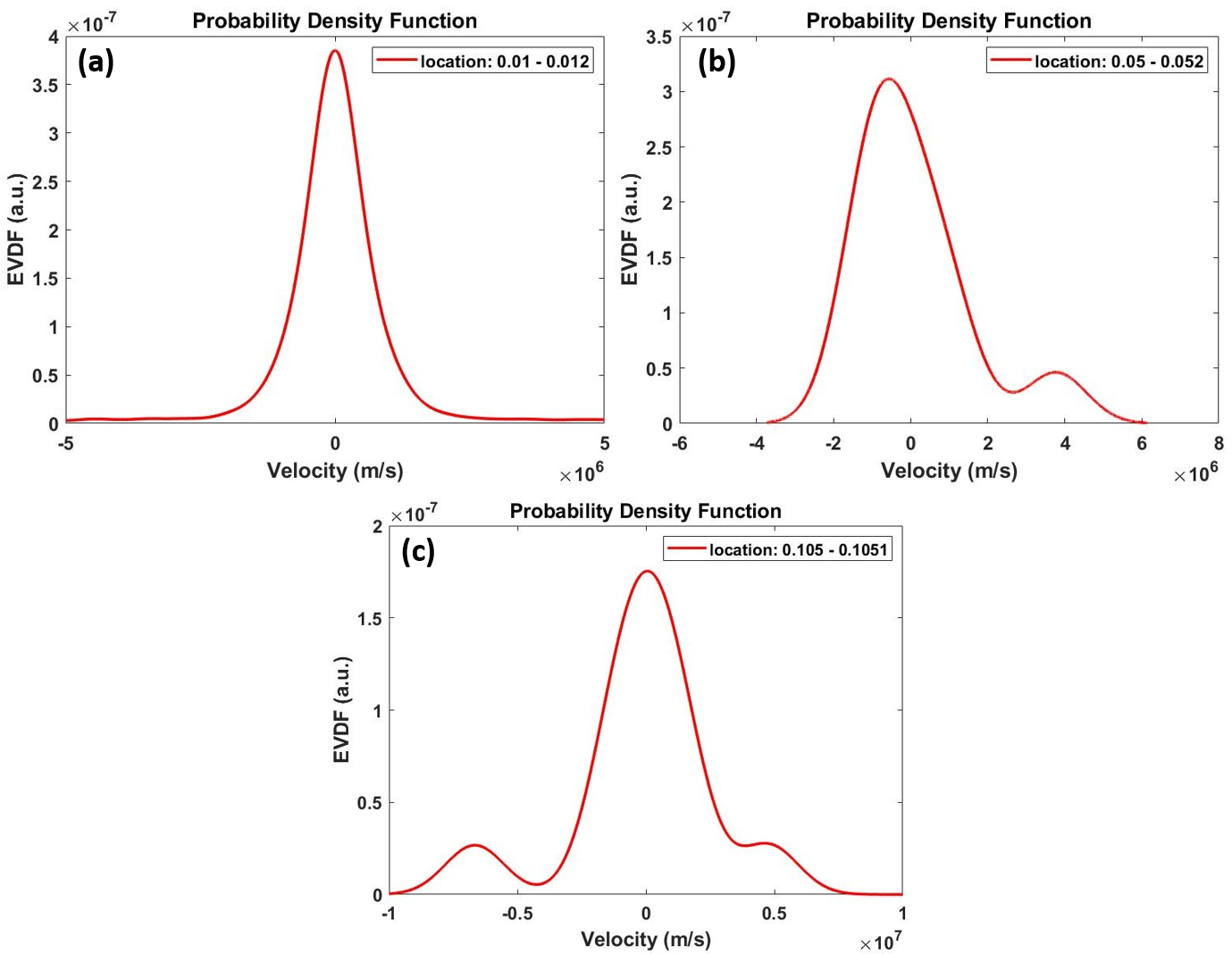}
	\caption{EVDF (a) near the electron emitter, (b) in-between emitter and the anode, and (c) inside the anode. Different locations can be visualised from the phase space plot [fig(\ref{fig:3})].\label{fig:5}}%
\end{figure}
Figure(\ref{fig:5}) shows the electron velocity distribution function (EVDF) at the prescribed locations. Near the wall region [fig(\ref{fig:5}{a})], the velocity distribution is observed to be Maxwellian. On the other hand, another distinct peak in the positive velocity axis is observed apart from the central peak in the second location [fig(\ref{fig:5}{b})]. The population of the electrons in this peak are those high energetic electrons which are moving in the forward direction towards the anode. Again, inside the anode region high velocity peaks are observed in both the sides of the central peak [fig(\ref{fig:5}{c})]. These peaks represent the electrons gained energy due to the high positive potential applied to the anode grids. The right peak in the positive velocity axis represents the population of the high velocity electrons moving in the forward direction. While, the peak in the left side of the central peak represents those electrons which are moving in the opposite direction. The velocity distribution of the electrons at different locations shows the re-circulatory characteristics of the energetic electrons across the anode grids. If we observe the EVDF in the second location [fig(\ref{fig:5}{b})], the broad nature of the Maxwellian distribution also suggests the presence of electrons moving in the backward direction.

\subsection{X-ray production, imaging and radiography}
One of the most primary objective of positive biasing to the central grid in the IECF system is the production of x-ray from the device and to perform x-ray imaging or radiography.
\begin{figure}
	\centering
	\includegraphics[width=8cm]{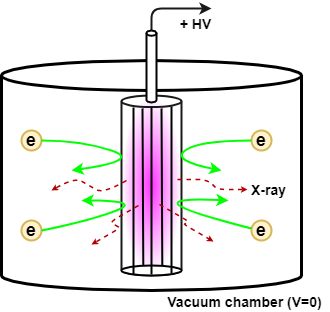}
	\caption{Conceptual picture of production of x-ray from the IECF device having a cylindrical anode grid.\label{fig:6}}%
\end{figure}
As already mentioned, the fast moving electrons interact with the anode grids due to the high positive voltage applied to the grids. In this interaction process the electrons rapidly decelerated by the target atoms. However, not every electrons decelerated in the same manner. Some of them are stopped in one impact and lose all their kinetic energy, while, some other deviates from its original path due to the target atoms and successively losing some of their kinetic energy. The electrons which are stopped in a single impact produces photons having maximum energy. The whole energy ($eV$) of the electron is converted into maximum photon energy ($h\nu _{max}$), where, $\nu _{max}$ is the maximum frequency of the photon. If an electron undergoes a glancing impact or deviated from its original path and partially loses its velocity, then only a fraction of its energy is converted into radiation and the photon thus produced has an energy less than the maximum value, $h\nu _{max}$. In the electron density profile, as shown in fig.(\ref{fig:4}), the formation of a high density electron cloud region is observed inside the anode grids. The process of deceleration of the electrons and hence the production of bremsstrahlung or continuous x-ray may also occurred due to the presence of the high density electron cloud inside the anode\cite{miley2014inertial}. The electron cloud takes part in the partial repulsion of the incoming high velocity electrons and may produce a fraction of bremmstrahlung radiation. Figure(\ref{fig:6}) shows the conceptual picture of generation of x-ray in the IECF device having a cylindrical anode grid at the center. \par
\begin{figure}
	\centering
	\includegraphics[width=8cm]{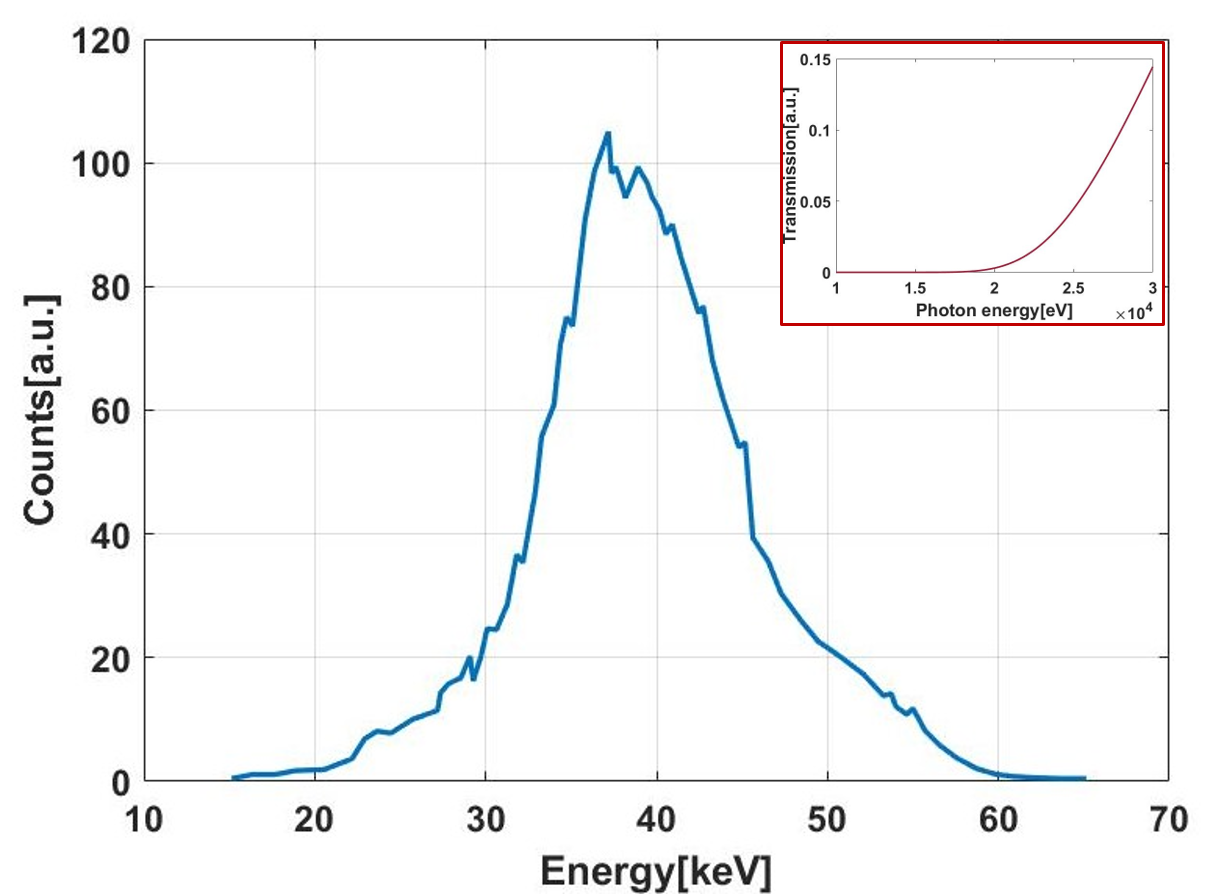}
	\caption{X-ray energy spectrum at $60$ kV anode potential. Transmission probability of the x-ray photon through the glass window is shown in the inset. \label{fig:7}}%
\end{figure}

 As discussed earlier in experimental set-up section, a PMT having NaI scintillator detector has been used to obtain the x-ray energy spectrum from the source. The detector is initially calibrated with a known gamma source of Cs-137. A typical spectrum at $60$ kV applied voltage and $5$ mA current is shown in fig.(\ref{fig:7}). It shows the continuous x-ray spectrum which extends up to the applied voltage range. It mostly resembles with the anode spectral model\cite{boone1997accurate}, having a linear decline towards the tail region. We have not observed any signature of characteristic radiation from the tungsten anode. However, application of high voltage up to $100$ kV may give rise to characteristics emissions, also. 

\subsubsection{X-ray imaging using pinhole camera}
\begin{figure}
	\centering
	\includegraphics[width=8cm]{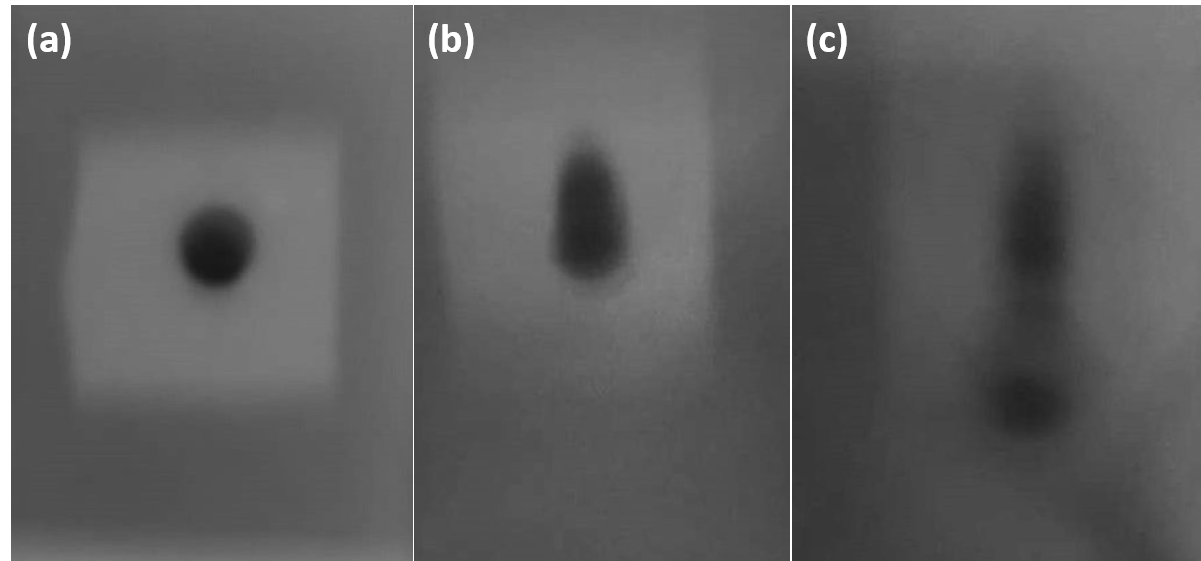}
	\caption{Pinhole images of the x-ray source using a spherical anode at an applied voltage of $35$ kV. Distance of the film is at (a) $0.1$ cm, (b) $1.0$ cm, and (c) $2.0$ cm from the pinhole.\label{fig:8}}%
\end{figure}

In order to observe the x-ray emitting zone we have used the pinhole camera and the x-ray film to record the image\cite{bhuyan2004comparative,bhuyan2011temporal}. The pinhole is fixed at the glass window from outside and the distance of the film is varied. At first, the spherical grid of $5$ cm diameter is used as the anode. The grid is attached to an extension rod so that it remains fixed at the central part of the chamber. After operating at an applied voltage of $35$ kV, $3$ mA current for $\sim 1$ minutes, the film is processed to get the final image. The image of the x-ray source region is shown in fig.(\ref{fig:8}{a}) when the film is at a distance of $0.1$ cm from the pinhole. 
The length of the x-ray source can be evaluated by using the relation\cite{steinmetz1982neutron}
\begin{equation}\label{eq:2}
    \beta = \frac{s^{/}}{s} = \frac{L^{/}}{L}
\end{equation}
Here, $\beta$ is the magnification of the image, $s$ and $s^{/}$ are the source and image distance from the pinhole, respectively, and $L$ and $L^{/}$ are the length of the source and image, respectively. The '$s$' is fixed at a distance of $35$ cm in all the cases. Now, if we increase the image distance, '$s^{/}$' (or film distance), then the magnification of the image also increases. Figure (\ref{fig:8}{b}) and (\ref{fig:8}{c}) shows the images of the source when the film is at $1.0$ cm and $2.0$ cm from the pinhole, respectively. The length of the source ($L$) is found to be $\sim 18$ cm by measuring the length of the images and using the eq.(\ref{eq:2}), in both the cases. Although, the total length of the spherical grid along with the extension rod is $25$ cm, since the part of the x-ray emitting from the upper portion of the anode is cut off by the area of the chamber in which the glass window is attached, the film is not able to collect the radiation from that part. Therefore, we are getting the remaining $18$ cm part of the source in the film. The structure or shape of the spherical anode grid can also be visualised in the lower part of the fig.(\ref{fig:8}{c}). \par We have replaced the spherical grid by an extension rod (solid cylindrical shape) and repeat the same experiment in order to analyse the x-ray source, in this case.
\begin{figure}
	\centering
	\includegraphics[width=8cm]{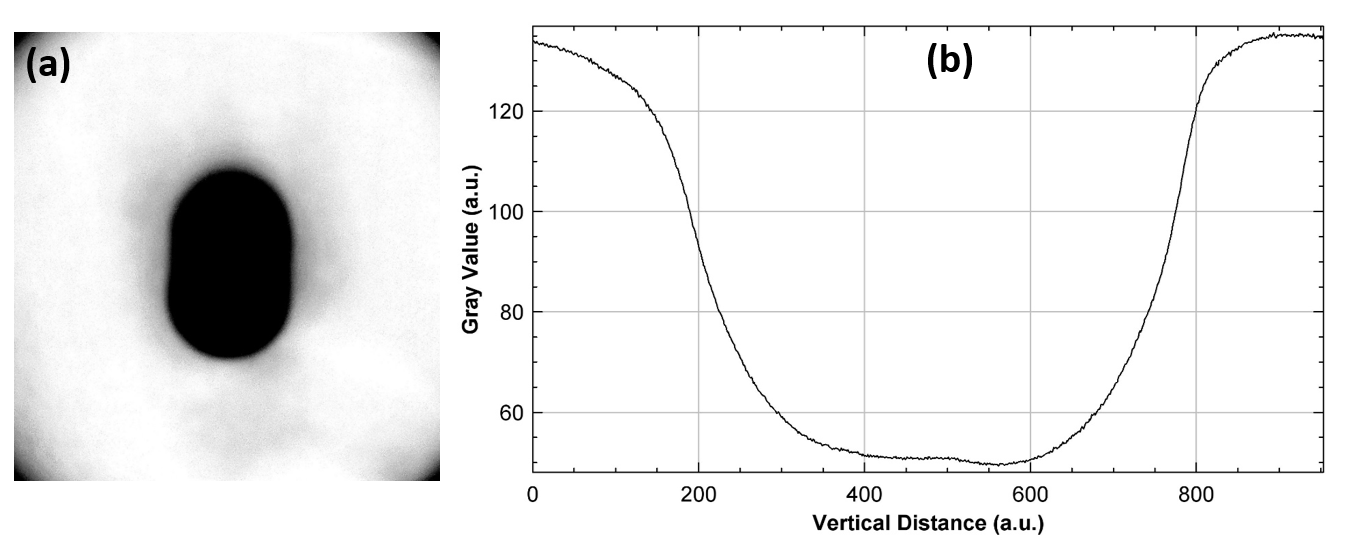}\\
	\includegraphics[width=8cm]{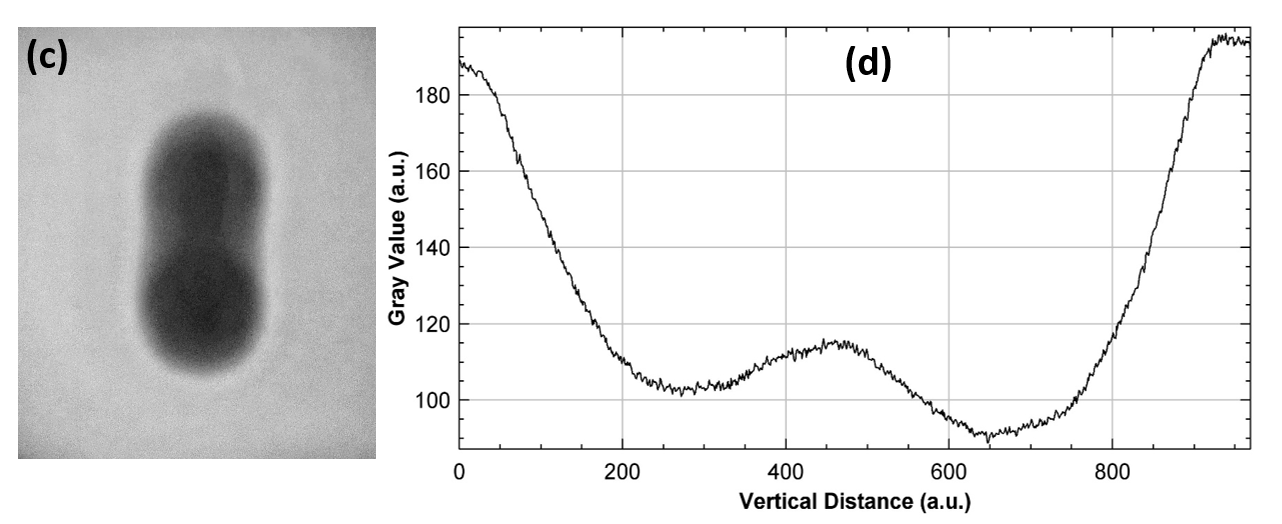}\\
	\includegraphics[width=8cm]{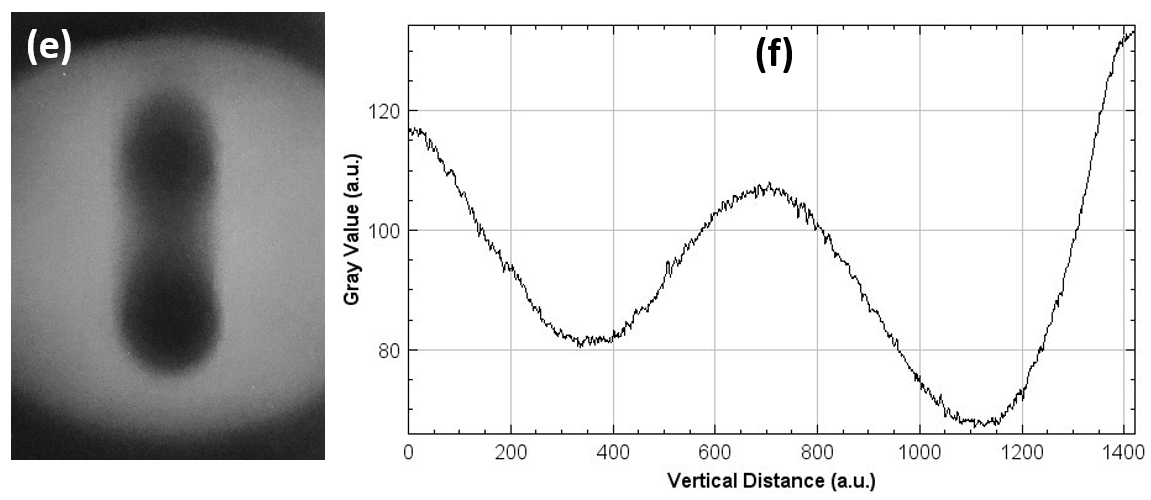}
	\caption{Pinhole images of the x-ray source using an extension rod as the anode at an applied voltage of $35$ kV. Distance of the film is at (a) $0.1$ cm, (b) $1.0$ cm, and (c) $2.0$ cm from the pinhole. The corresponding gray scale intensity levels in the vertical direction of the images are shown in (b), (d), and (f), respectively. \label{fig:9}}%
\end{figure}
The results are shown in fig.(\ref{fig:9}) with the corresponding gray scale intensity level of the images in the vertical direction. It has been noticed that the intensity in the central part of the images are decreased and it becomes prominent with the increase in magnification of the image. The primary reason behind it may be the formation of asymmetric electric field in the middle part of the device due to the presence of multiple number of ports, diagnostic tools, gas inlet and outlet, etc. in that region. The motion of the electrons will be asymmetric in the mid section in comparison to the upper and lower sections of the device. As a result, the variation in intensity are observed in all the images. Similar variation is also observed in the case of spherical anode [fig.(\ref{fig:8}{c})]. \par Lastly, the cylindrical grid is used as the anode and similar type of intensity variation of the images have been noticed. This time, we have tried to observe the images of the source by putting the pinhole arrangement along with the film inside the chamber (attached the arrangement on the glass window from inside). However, extra precautionary measures have been taken in order to keep the x-ray film safe from heat and light inside the chamber. 
\begin{figure}
	\centering
	\includegraphics[width=8cm]{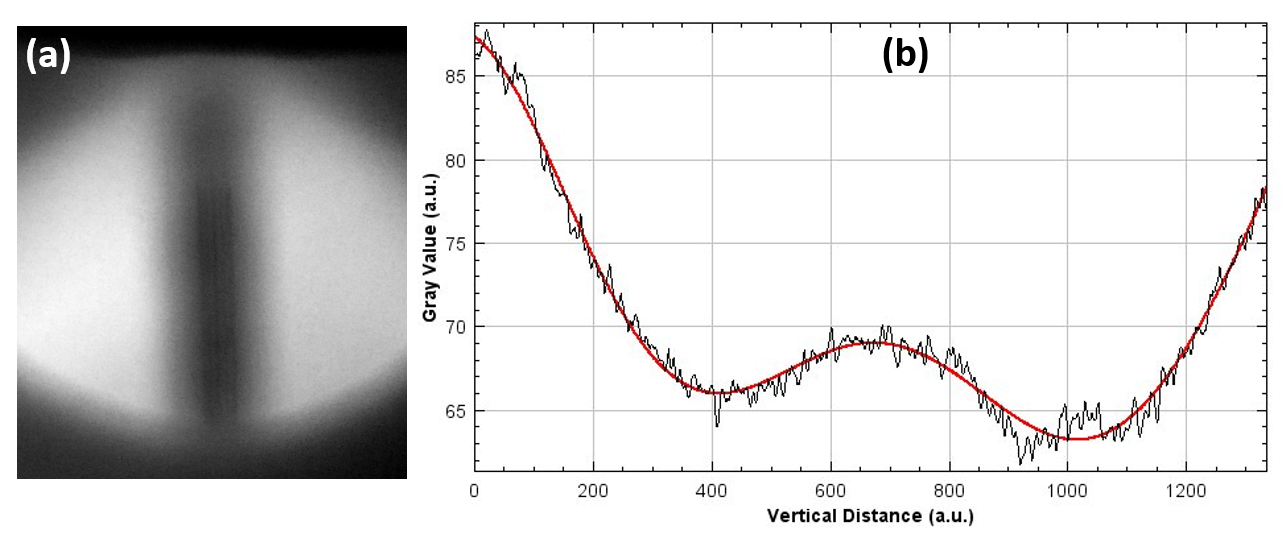}\\
	\includegraphics[width=8cm]{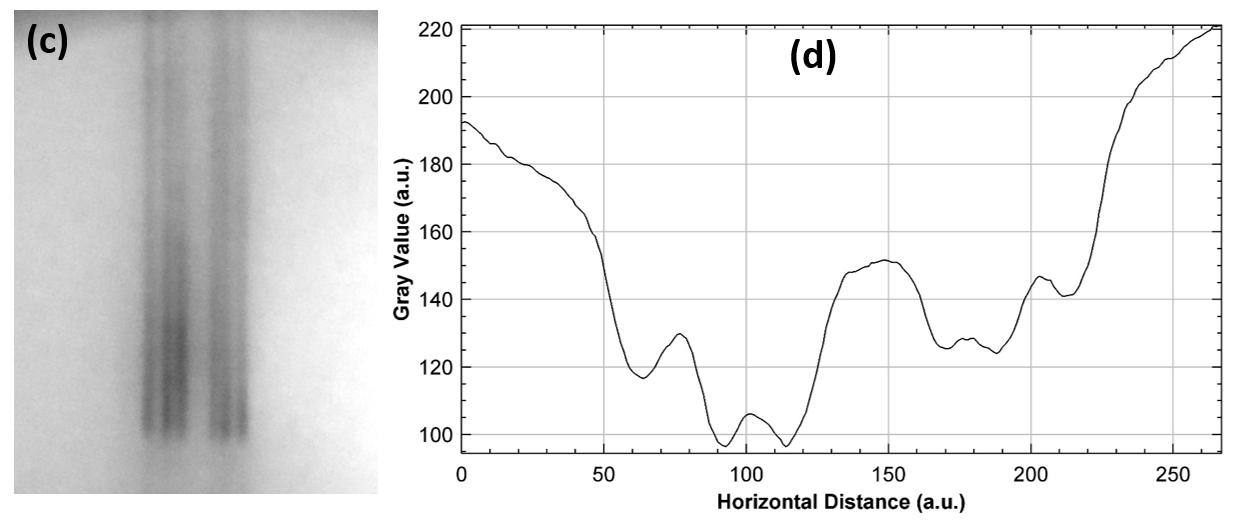}
	\caption{Pinhole images of the x-ray source using the cylindrical anode having (a) $12$, and (c) $6$ number of grid wires. Pinhole arrangement is attached to the glass window from inside the chamber. Corresponding gray scale intensity level of (a) in the vertical direction is shown in (b), and that of (c) is shown in (d) in the horizontal direction. The red line is the best fitted line. \label{fig:10}}%
\end{figure}
After, operating at $35$ kV for $1$ minute the film has been processed and the results are shown in fig.(\ref{fig:10}). In this case, two grids having $12$ and $6$ number of grid wires have been used, separately. The hazy image of the grid wires of the anode having $12$ number of grids have been noticed if closely observed, as shown in fig.(\ref{fig:10}{a}). Since, the grid wires are so closely packed the clear image of the grids are difficult to obtain, unless the magnification of the image is increased. However, with the increase in magnification the resolution of the image decreases. The corresponding gray level of the intensity of the image in the vertical direction is shown in fig.(\ref{fig:10}{b}). It shows the similar type of intensity variation, with a slight decrease of intensity in the mid section, as obtained in previous cases. In case of the $6$ gridded cylindrical anode, the clear image of the grid wires can be observed, as shown in fig.(\ref{fig:10}{c}). The six distinct troughs in the corresponding gray scale intensity profile [fig.(\ref{fig:10}{d})] in the horizontal direction of the image signifies the six grid wires. The hot spot or the exact x-ray emitting zone have been perfectly observed, in this particular case. It has also been noticed that such clear images of the grid wires are not obtained outside the chamber. The absence of any dark region inside the grid also discards any contribution of the electron cloud in x-ray formation. In fact, the six gridded anode unable to form a highly dense electron cloud inside it which might be the reason for not getting the x-ray from that region, as shown in fig.(\ref{fig:10}{c}).

\subsubsection{X-ray radiography}
\begin{figure}
	\centering
	\includegraphics[width=8cm]{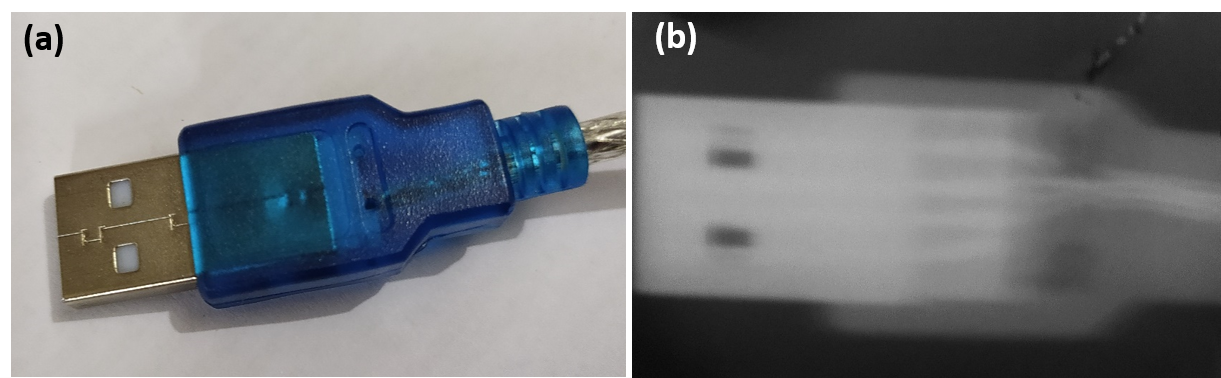}\\
	\includegraphics[width=8cm]{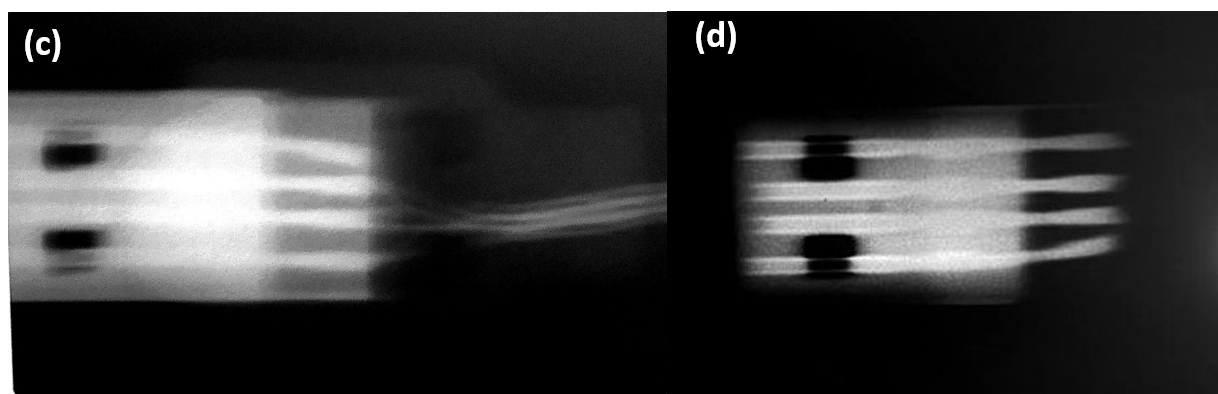}
	\caption{X-ray radiography of a USB plug. (a) The actual photograph. Radiography images at (b) $25$ kV, (c) $35$ kV, and (d) $45$ kV applied voltages showing the dependence of applied voltage on the resolution of the image formed. Anode used is the cylindrical grid having $12$ number of grid wires. \label{fig:11}}%
\end{figure}

As far as x-ray radiography is concerned, we have placed the sample objects on the glass window from outside and also attached the x-ray film on it. The resolution and quality of the image depends on the anode apart from the applied voltage and current. Since, we have used gridded structure as the anode, its transparency or the number of grid wire present in the anode plays an important factor upon which the quality of the image depends. The image quality and resolution using a $12$ gridded cylindrical anode, e.g., is much better than that of a $6$ gridded anode having the same diameter and applied voltage. Again, the dependency of applied voltage on the radiography images is shown in fig.(\ref{fig:11}). Here, we have used a USB plug as the object to show the variation of its radiography images by varying the applied voltage (constant current: $10$ mA). As we increase the voltage, the interior parts of the sample becomes more prominent, gradually. At $45$ kV applied voltage, the outer part of the USB plug along-with the connecting wires become invisible and only the hard metallic part of it remains in the image, as shown in fig.(\ref{fig:11}{d}).  
\par
\begin{figure}
	\centering
	\includegraphics[width=8cm]{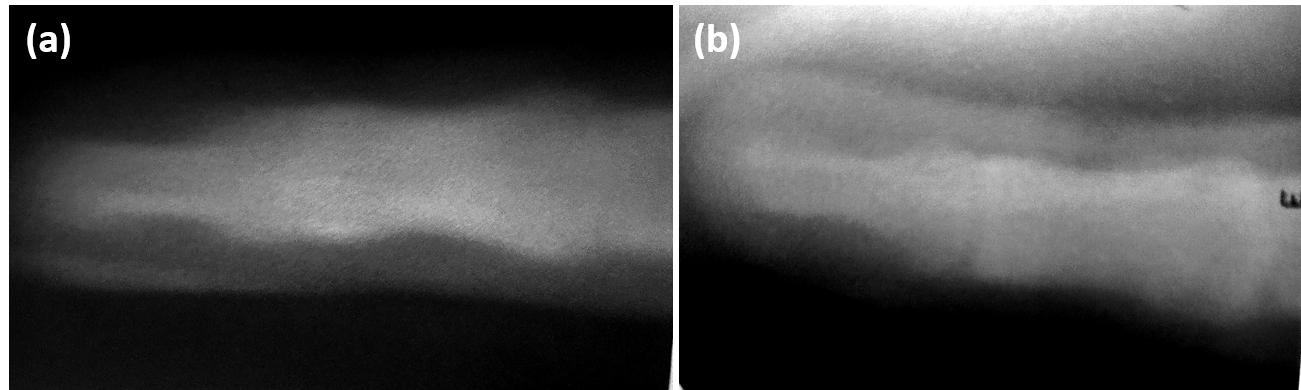}
	\caption{Radiography image of the human little finger at $35$ kV applied voltage using (a) cylindrical grid of $2$ cm diameter and having $12$ number of grid wires, and (b) single extension rod of $1.5$ cm diameter.\label{fig:12}}%
\end{figure}

 On the contrary, when we replaced the cylindrical grid by the solid extension rod then also we can observe the difference in the image quality. One such example is shown in fig.(\ref{fig:12}), which shows the radiography image of the human little finger at $35$ kV applied voltage. The bone structure of the finger can be visualised from the image using the cylindrical grid [fig.(\ref{fig:12}{a})], while, the image obtained using the solid rod as the anode [fig.(\ref{fig:12}{b})] displays some more detailed information, such as, the knuckles of the finger can be observed with higher contrast level. The optimisation of the applied voltage is one of the key parameters while taking the radiography image of a biological sample. If the voltage is too low (less than $25$ kV), only the flesh part of the finger appears without any visualisation of the bone. Again, if it is too high (more than $50$ kV) then x-ray even passes through the bone and the structure looks to be darker. Moreover, the expose duration of the sample to the radiation is also equally important for getting clear image. It has been observed that $50-60$ seconds is the appropriate exposure time to obtain images, as shown in fig.(\ref{fig:12}). Longer or shorter exposure time results unclear images of the sample (finger).

\section{Conclusion}
This study showcased the applicability of the IECF device as an x-ray source by simply reverting the polarity of the central grid. The distinct features of this study includes the computational observation of the electron characteristics through PIC simulation. It shows the re-circulating nature of the electrons across the grid. The high energy peaks in the EVDF signify the high velocity electrons moving in the forward and backward direction through the grid. The high density electron cloud formed inside the anode contributes in the x-ray production apart from the contribution of electron-anode interaction. The detected continuous or bremsstrahlung radiation shows a broad energy spectra extending up to the applied voltage. The pinhole images of the x-ray source shows a slight decrease in image intensity in the middle portion. The asymmetric field distribution in the mid portion due to the presence of multiple number of ports might be the reason. Lastly, both metallic and biological samples are showing good radiography images depending upon the applied voltage, current and the structure of the anode. \par The table-top and small-scale neutron/x-ray sources, such as the IECF is always desirable which has the capability to replace the large-scale accelerator-based devices in the near future. The IECF device may become an alternative when both neutron and x-ray scanning facilities are needed for greater accuracy in the security system.

\begin{acknowledgments}

The authors are grateful to the Director, Institute for Plasma Research (IPR), Gandhinagar, India and the Center Director, Center of Plasma Physics-Institute for Plasma Research (CPP-IPR), Sonapur, India, for providing us the opportunity to carry out this work. We are also thankful to Mr. M.K.D. Sarma for his technical support.
\end{acknowledgments}

\section*{Data Availability Statement}
The data that supports the findings of this study are available within the article.

\nocite{*}
\bibliography{aipsamp}

\end{document}